\documentclass[preprint,12pt]{aastex}

\shorttitle{Molecular outflows in Massive Young Stars} \shortauthors{Qiu et al.}

\begin{document}

\title{High Resolution Imaging of Molecular Outflows in Massive Young Stars}

\author{Keping Qiu}
\affil{Harvard-Smithsonian Center for Astrophysics, 60 Garden Street, Cambridge MA 02138, USA \\ Department of
Astronomy, Nanjing University, Nanjing 210093, China} \email{kqiu@cfa.harvard.edu}

\author{Qizhou Zhang}
\affil{Harvard-Smithsonian Center for Astrophysics, 60 Garden Street, Cambridge MA 02138, USA}

\author{Henrik Beuther}
\affil{Max-Planck-Institute for Astronomy, K\"onigstuhl 17, 69117 Heidelberg, Germany}

\and

\author{Ji Yang} \affil{Purple Mountain Observatory, Chinese Academy of Sciences, Nanjing 210008, China}

\begin{abstract}
We present high angular resolution observations toward two massive star forming regions IRAS 18264-1152 and IRAS
23151+5912 with the Plateau de Bure Interferometer (PdBI) in the SiO (J=2-1) and H$^{13}$CO$^{+}$ (J=1-0) lines
and at 1.3 mm and 3.4 mm continuum, and with the Very Large Array (VLA) in the NH$_3$ (J,K)=(1,1), (2,2) lines.
The NH$_3$ (1,1) and (2,2) emission is detected toward IRAS 18264-1152 only. For IRAS 18264-1152, the SiO
observations reveal at least two quasi-perpendicular outflows with high collimation factors, and the most dominant
feature is a redshifted jet-like outflow with very high velocities up to about ${\Delta}v=60$ kms$^{-1}$ with
respect to the systemic velocity. The very-high-velocity component (${\Delta}v=22 - 60$ kms$^{-1}$) of this
outflow is spatially offset from its high-velocity (${\Delta}v=3 - 21$ kms$^{-1}$) component. The SiO line
profiles and position-velocity characteristics of these two components suggest that this outflow can be driven by
an underlying precessing jet. For IRAS 23151+5912, the bipolar but mainly blueshifted SiO outflow coincides with
the inner parts of the single-dish CO outflow. In particular the quasi-parabolic shape of the blueshifted outflow
coincides with the near-infrared nebulosity and is consistent with entrainment of the gas by an underlying
wide-angle wind. The analysis of the molecular outflow data of the two luminous sources further support high-mass
stars forming via a disk-mediated accretion process similar to low-mass stars.

\end{abstract}


\section{Introduction}
The formation process of massive stars is poorly understood. There are two competing views: One view is that
high-mass stars form via infall and disk mediated accretion processes as a scaled-up version of their low-mass
counterparts \citep[e.g.,][]{Jijina96, Garay99, Norberg00, McKee02, Yorke02, McKee03, Keto03}. The other view
argues that at the dense center of evolving massive clusters, competitive accretion could become a dominant
process. In the extreme case, even the coalescence of low- and intermediate-mass (proto)stars may occur to form
the most massive stellar objects \citep[e.g.,][]{Bonnell98, Stahler00,Bonnell04, Bally05}.

Massive molecular outflows play a very important role in massive star formation. In addition to dissipating excess
angular momentum of the infalling material as their low-mass counterparts \citep{Shu87}, they may inject enough
energy to sustain the turbulence and prevent a decrease of the virial parameter \citep{Quillen05}, whose value may
determine whether competitive accretion is effective in cluster forming clouds \citep{Krumholz05a}. Outflows also
lead to significant anisotropy in the distribution of the circumstellar material, which greatly reduces the
radiation pressure experienced by the infalling material \citep{Krumholz05b}. The collimation factor of massive
molecular outflows can help to discriminate between the disk mediated accretion process and the coalescence
process, since the former requires massive outflows as collimated as their low-mass counterparts and the latter
predicts outflows to be far less collimated.

Surveys with single-dish telescopes find that massive molecular outflows are ubiquitous toward high-mass star
forming regions, and are far more massive and energetic than their low-mass counterparts
\citep[e.g.,][]{Shepherd96, Zhang01, Beuther02c, Zhang05}. While \citet{Shepherd96} find that massive outflows
appear to be less collimated than their low-mass counterparts, \citet{Beuther02c} argues that this may be mostly
an observational artifact caused by the large distances of the target sources and the low spatial resolution of
most studies. In recent years, there have been several high spatial resolution mm interferometric studies of
molecular outflows toward massive star forming regions. \citet{Shepherd99} reveal a massive outflow with a wide
opening angle toward G192.16-3.82 and suggest its driven mechanism as a strong wide-angle wind; \citet{Cesaroni99}
report a highly collimated SiO jet toward IRAS 20126+4104, and \citet{Shepherd00} reveal the same, but a larger
scale (2pc) outflow in CO in a different orientation; \citet{Hunter99} show a well collimated SiO jet toward AFGL
5142; Beuther et al. (IRAS 05358+3543, 2002a; IRAS 19410+2336, 2003; IRAS 19217+1651, 2004) resolve simple
outflows previously identified with single-dish observations into multiple and well collimated outflows when
observed with interferometers at high spatial resolution. They find that their kinematics are similar to those of
low-mass counterparts; \citet{Shepherd03} present multiple, overlapping massive outflows driven by at least four
protostars in the region of W75\,N. Their energetics and near-infrared observations suggest that they are not
likely to be scaled-up versions of jet-driven outflows from low-mass protostars; \citet{Gibb03} show that a
precessing jet-driven flow associated with G35.2-0.7N in single-dish observations can be explained as at least two
overlapping flows; \citet{Su04} reveal a clear bipolar morphology of molecular outflows toward two luminous
sources IRAS 21519+5613 and IRAS 22506+5944; \citet{Kumar04} unveil at least four outflows toward an ultra-compact
(UC) H{\footnotesize II} region Onsala 1; \citet{Sollins04} resolve a highly energetic bipolar SiO outflow toward
UC H{\footnotesize II} region G5.89-0.39. All of these interferometric observations reveal new morphologies of
massive outflows that can not be resolved with single-dish observations and give a wealth of information about the
physical process in the innermost parts of the massive star forming regions.

As the statistics of the high spatial resolution interferometric observations are still poor and some basic
issues, such as collimation factors and outflow driving mechanism, are being debated, we carry out a study of
massive molecular outflows with the Plateau de Bure Interferometer (PdBI) and the Very Larege Array (VLA). Here we
use the shock tracer SiO (J=2-1) to image outflows toward two massive star forming regions IRAS 18264-1152 and
IRAS 23151+5912 (hereafter we refer as I18264 and I23151, respectively). The dense cores and ambient gas are
mapped at 1.3 and 3.4 mm continuum, and in the NH$_3$(1,1) and (2,2) and H$^{13}$CO$^+$ (J=1-0) lines. The two
sources are part of a larger sample discussed in detail by \citet{Sridharan02} and \citet{Beuther02b, Beuther02d}.
\citet{Beuther02c} present bipolar CO outflows toward the two sources with the IRAM 30m telescope observations.
Toward I23151, \citet{Weigelt06} reveal a cone-like nebulosity, which coincides with the blueshifted CO outflow,
in the near-infrared $K^{'}$ band using the bispectrum speckle interferometry method. The kinematic distance of
I23151 derived from the CS velocity is 5.7 kpc. The kinematic distance of I18264, suffering from an ambiguity, is
either 3.5 or 12.5 kpc \citep{Sridharan02}. By associating the region via the near- and mid-infrared surveys 2MASS
and MSX on larger scales with sources of known distance, the ambiguity was solved to be at the near distance (S.
Bontemps, 2003, priv com.). On the other hand, toward the line of sight of I18264, \citet{Sewilo04} solve the
kinematic distance ambiguity of an H{\footnotesize II} region with velocity of 50.9 kms$^{-1}$ to be at the far
distance of 12.0 kpc. For the molecular cloud with velocity of 43.6 kms$^{-1}$, Sewilo et al. consider it is
associated with the H{\footnotesize II} region of 50.9 kms$^{-1}$ and adopt its far distance of 12.4 kpc. Since
the distance determinations above both use indirect methods and still give ambiguous results, we give physical
parameter estimations corresponding to both the far and near distances in the following sections. Based on the
High Resolution (HIRES) IRAS database, \citet{Sridharan02} estimate the luminosities of the two sources to be
$10^5 L_{\odot}$ for I23151 and $10^4 L_{\odot}$ and $1.2\times10^5 L_{\odot}$ respectively for the near and far
distances for I18264. As additional evidence of massive star formation and outflows, Class {\footnotesize II}
CH$_3$OH and H$_2$O maser emission toward I18264 and H$_2$O maser emission toward I23151 is observed
\citep{Beuther02d}. \citet{Sridharan02} detect no emission at 3.6 cm down to 1 mJy for both sources, while
\citet{Zapata06} detect 1.3 cm and 3.6 cm emission toward I18264 with a better sensitivity. After a description of
observations in Sect. \ref{obs}, we show results in Sect. \ref{result}. Discussion is presented in Sect.
\ref{dis}, and we conclude in Sect. \ref{conclusion}.

\section{Observations} \label{obs}
The observations of I18264 and I23151 were carried out during August 2003 to November 2003 using the PdBI
\footnote{Based on observations carried out with the IRAM Plateau de Bure Interferometer. IRAM is supported by
INSU/CNRS (France), MPG (Germany) and IGN (Spain).}. At the 3mm wave band, the spectral correlator was set to
sample the SiO (v=0, J=2-1) and H$^{13}$CO$^{+}$ (J=1-0) lines with a bandwidth of 40 MHz. In addition, two wide
bands of 320 MHz were set to observe the continuum emission, and to cover the SiO and H$^{13}$CO$^+$ lines as
well. Taking advantage of the dual frequency operation, we set the correlator to the CN (J=2-1) line using a
bandwidth of 40 MHz, and two 320 MHz wide bands at the 1mm band for continuum. The CN (2-1) emission was not
detected. Except for observations toward I23151 on 2003 December 9, all the tracks were carried out in weather
conditions adequate for the 3mm band only. I18264 was covered with a 7-field mosaic with the pointing centers at
$({\Delta}RA, {\Delta}Dec) = (5'', 3''\!.5; 28'', 3''\!.5; -18'', 3''\!.5; 16''\!.5, 23''\!.5; -6''\!.5, 23''\!.5;
16''\!.5, -16''\!.5; \\-6''\!.5, -16''\!.5)$ with respect to the reference center at RA(J2000) = 18:29:14.3 and
Dec(J2000) = -11:50:26. I23151 was observed with a single field with the pointing center at RA (J2000) =
23:17:21.0 and Dec (J2000) = 59:28:49.00. We used MWC349 as the primary flux calibrator, and 3C273/3C345 as the
bandpass calibrators. The time dependence gain was monitored by observing 1741-038, 1830-210, and 2200+420. The
visibility data were calibrated and imaged using the standard procedure in GILDAS package. The systemic velocities
are 43.6 kms$^{-1}$ for I18264 and -54.4 kms$^{-1}$ for I23151. The nominal spectral resolution was about 0.4
kms$^{-1}$ and averaged to about 1 kms$^{-1}$ for SiO (2-1). The continuum rms in the 3mm band was $\sim$0.7 mJy
for I18264 and $\sim$0.2 mJy for I23151.

The observations of the NH$_3$ (1,1) and (2,2) lines were carried out on 1997 December 13 for I23151 in the D
configuration, and on 2001 July 23 for I18264 in the C configuration of the VLA \footnote{The National Radio
Astronomy Observatory is operated by Associated Universities, Inc., under cooperative agreement with the National
Science Foundation.}. In both observations, we used the correlator mode 4 which provided 3.13 MHz bandwidth for
both the left and right polarizations of the (1,1) and (2,2) lines, respectively. The spectral resolutions of the
observations was 48 KHz, or 0.6 kms$^{-1}$ at the NH$_3$ line frequencies. The pointing center of the observations
was RA(J2000) = 18:29:14.31, Dec(J2000) = -11:50:25.6 for I18264 and RA(J2000) = 23:17:21.10, Dec(J2000) =
59:28:48.6 for I23151. The on-source integration per source was about 1 hour. The visibility data were calibrated
and imaged in the AIPS package. The left and right polarizations were averaged during imaging to reduce the noise
level in the data. The synthesized beam size was about 4$''$ for the I23151 images, and $2''\!.8 \times 2''\!.0$
for I18264 images. The rms noise level is 7 mJy per 0.6 kms$^{-1}$ channel.

\section{Results} \label{result}
For both sources, we detected strong emission in SiO (J=2-1) and H$^{13}$CO$^+$ (J=1-0). I18264 has detectable
ammonia emission at an rms of 7 mJy per 0.6 kms$^{-1}$ channel, and I23151 was not detected in NH$_3$ at an rms of
8 mJy per 0.6 kms$^{-1}$ channel.

\subsection{Millimeter Continuum sources}
\begin{figure}
\plotone{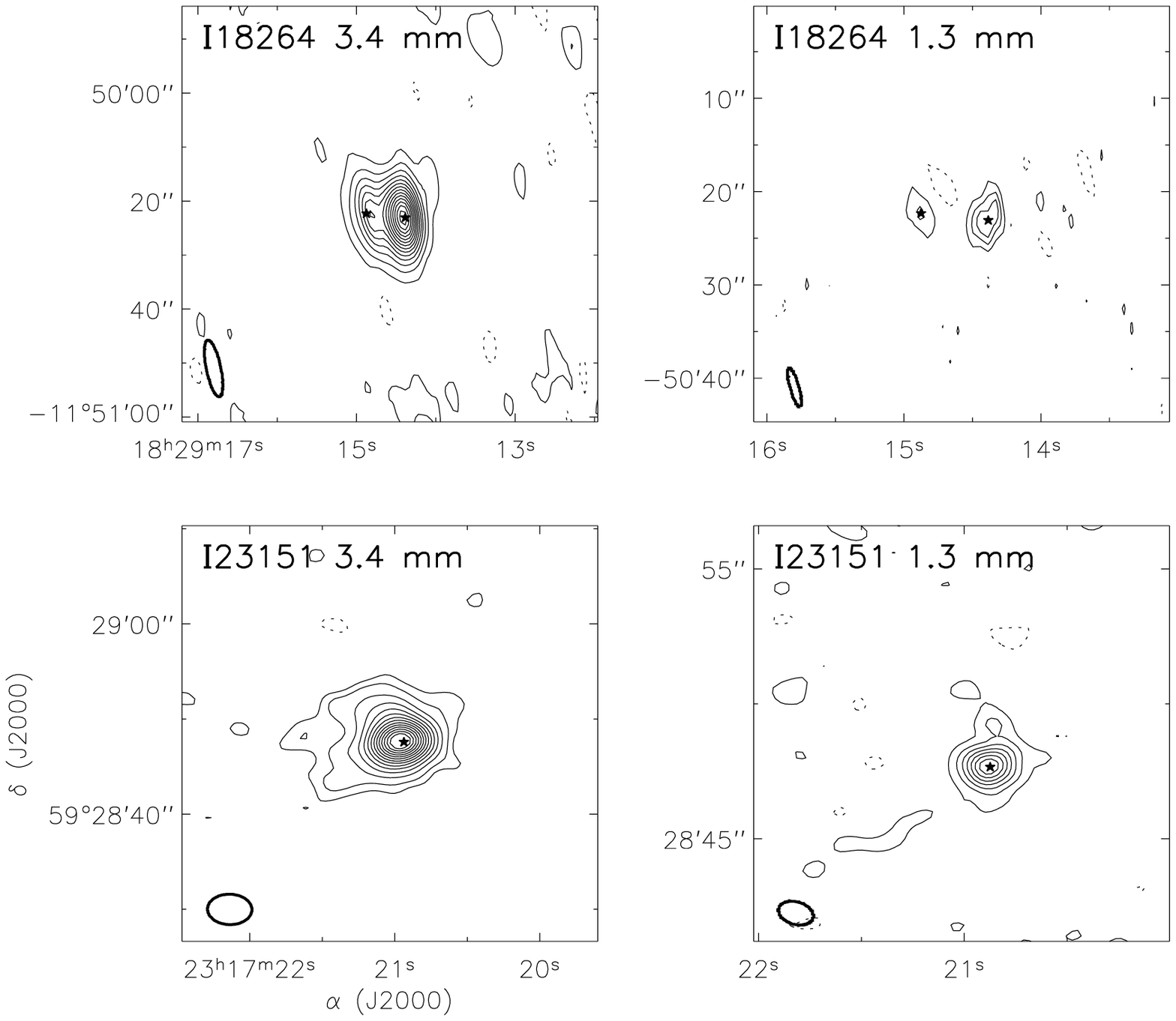} \caption{\textbf{Top:} the 3.4 mm ({\it left}) and 1.3 mm ({\it right}) continuum emission in
I18264. The lowest and spacing contours are 2.1 mJy ($3\sigma$) at 3.4 mm and 22.5 mJy ($3\sigma$) at 1.3 mm. The
dashed contours represent the $-3\sigma$ level. \textbf{Bottom:} similar to the top but for I23151. The lowest and
spacing contours at 3.4 mm are 0.6 mJy ($3\sigma$).  At 1.3 mm, the contours start at 4.25 mJy ($2.5\sigma$) in
steps of 5.1 mJy ($3\sigma$). The dashed contours represent the $-3\sigma$ level for 3.4 mm and $-2.5\sigma$ level
for 1.3 mm. The stars hereafter mark the 1.3 mm continuum peaks. The beams are shown as thick ellipses at the
lower left of each panel. \label{cont}}
\end{figure}

Fig. \ref{cont} presents the 3.4 mm and 1.3 mm continuum emission of the two sources. For I18264, the 1.3 mm and
3.4 mm continuum emission resolve two peaks with the stronger one in the west. While for I23151, the continuum
emission remains singly peaked even at a resolution of $1''\!.3\times0''\!.85$ at 1.3 mm. For both sources, the
peak positions at 1.3 mm and 3.4 mm coincide well with each other. In I23151, a fan-shaped structure opening to
the east can be roughly identified in the 3.4 mm continuum emission. In the 1.3 mm continuum emission, there is a
discontinuous arch structure at the lowest contour level which goes from a little bit northeast to the southeast
of the mm continuum peak. We will further discuss these features in Sect. \ref{sio}. Because the weather
conditions were only adequate for the 3 mm wave band during the observing seasons, we use the 3.4 mm continuum
data for quantitative analysis. Assuming that the 3.4 mm continuum is mainly produced by optically thin dust
emission, we can calculate the masses of the dense cores following the relation
$M_{gas+dust}=F_{\nu}D^2/B_{\nu}(T_d)\kappa{_\nu}$ \citep{Hildebrand83}, where $F_\nu$ is the flux density of the
dust emission, D is the distance to the source, and $B_\nu$ is the Plank function at a dust temperature of $T_d$.
\citet{Sridharan02} derive dust temperatures of 35 K for I18264 and 68 K for I23151 by graybody fits to the IRAS
and mm data. Here the dust opacity per gram is taken to be $\kappa_\nu=0.1(\nu/1.2$THz$)^{\beta}$cm$^2$g$^{-1}$
\citep{Hildebrand83}, where the opacity index $\beta$ is set to be 1.5. The results of the calculations
($M_{core}$) are listed in Table. \ref{table1}. The uncertainty of this estimation mainly comes from the
determinations of $\beta$ and $T_d$. The masses will decrease by a factor of 4 if $\beta=1$, and increase by a
factor of 2 if $T_d$ decreases to $T_d/2$. The integrated flux at 3.4 mm is 0.13 Jy for I18264 and 0.029 Jy for
I23151. When compared with the 1.2 mm single-dish flux \citep{Beuther02a}, the total flux from PdBI amounts to
63\% for I18264 and 56\% for I23151 of the single-dish flux extrapolated from 1.2 mm using
$S({\nu})\,{\propto}\,{\nu}^{2+{\beta}}$ with $\beta=1.5$. Some extended emission is not recovered by the
interferometer.

\subsection{SiO outflows} \label{sio}
\begin{figure*}
\plotone{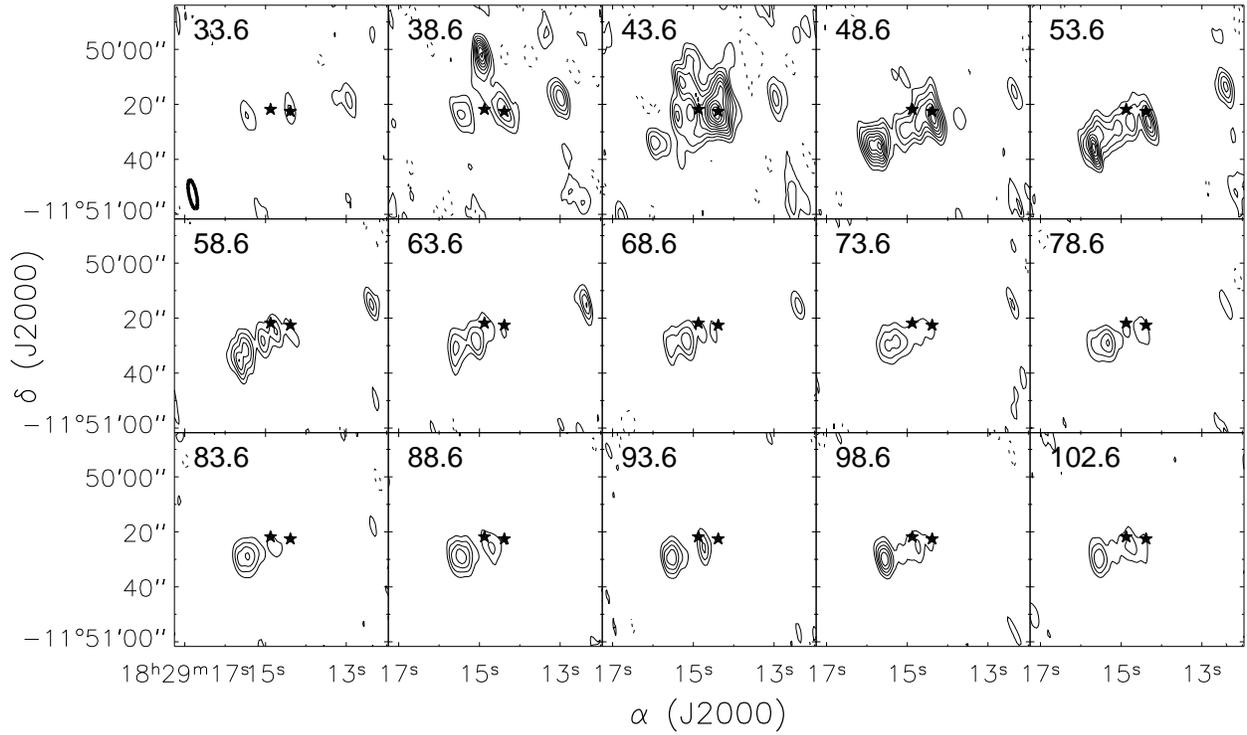} \caption{The SiO channel maps in I18264, with a smoothed velocity resolution of 5 kms$^{-1}$. The
contours start at 33 mJy (3$\sigma$) in steps of 22 mJy (2$\sigma$). The dashed contours represent the -3$\sigma$
level. The numbers in the top left of each panel denote the central velocity of each channel. The thick ellipse in
the lower left of the first panel shows the beam size. \label{siochan1}}
\end{figure*}

Fig. \ref{siochan1} presents the channel maps in SiO (2-1) in I18264, where the velocity resolution is smoothed to
5 kms$^{-1}$. The SiO emission mostly appears in the redshifted channels in Fig. \ref{siochan1}, and only the 38.6
kms$^{-1}$ channel shows prominent blueshifted emission. The most remarkable feature in the channel maps is an
elongated structure in the southeast. This redshifted emission has very high velocities up to $\Delta{v}\sim$60
kms$^{-1}$ with respect to the systemic velocity ($v_{LSR}$) 43.6 kms$^{-1}$. We have examined the data from the
320 MHz band and found that there is no detectable SiO emission beyond $v_{LSR}$ of 110 kms$^{-1}$. The integrated
blue- and redshifted SiO emission is shown in Fig. \ref{sioint1}a, where the single-dish bipolar CO outflow is
resolved into two quasi-perpendicular outflows: The southeast to northwest (SE-NW) outflow and the northeast (NE)
outflow. Both outflows seem to originate from the western peak of the mm continuum. Along the SE-NW outflow, both
red- and blueshifted emission can be found, which is a typical feature for expanding bow shocks near the plane of
the sky. From its alignment with the SE jet-like outflow, the bipolar emission in the northwest seems to be the NW
lobe of the SE-NW outflow. But it is also possible that this feature is due to another low-mass (proto)star whose
mass is below our detection limit. The two outflows are both well collimated with overall collimation factors
$\sim3$ for the NE outflow and $\sim4$ for the SE-NW outflow. The estimated collimation factors should be the
lower limits considering the unknown inclination angles and the spatial resolution limit. In Fig. \ref{sioint1}b,
the redshifted SiO emission is shown in two velocity ranges, i.e. the high-velocity component (HC) with velocities
${\Delta}v = 3 - 21$ kms$^{-1}$, and the very-high-velocity component (VHC) with velocities ${\Delta}v = 22 - 60$
kms$^{-1}$. The definition of the velocity ranges here is based on the morphological changes in the SiO emission
in the velocity channel maps and the characteristics in the position-velocity and mass-velocity diagrams that will
be discussed below. The redshifted SE outflow in the VHC is shifted toward north and its remote peak is closer to
the driving source than that in the HC. We will discuss this jet-like outflow in detail in Sect. \ref{kin}. Thanks
to the continuous velocity structures in this dominant outflow, we derive the dynamical time following
$$t_{dyn}=\frac{L_{flow}}{{\Delta}v_{max}},$$ where $L_{flow}$ is the length of the jet-like SE outflow and
${\Delta}v_{max}$ is the maximum velocity of the SE outflow. Assuming the optically thin thermal SiO (2-1)
emission in local thermodynamic equilibrium (LTE), we estimate the gas mass in the outflow according to
$$N_{SiO}=\frac{3k^2c^2}{2{\pi}^4h{\mu_d}^2{\nu}^4}\frac{T_{ex}}{\Delta\Omega}\,exp(\frac{E_J+h\nu}{kT_{ex}}){\int}S_{\nu}\,dv;$$
$$M_{out}=N_{SiO}\,[\frac{H_2}{SiO}]\,\mu_g\,m_{H_2}d^2\,{\Delta}{\Omega},$$
where $\mu_d$ is the permanent dipole moment, $\Delta\Omega$ is the FWHM of the synthesized beam, $\mu_g=1.36$ is
the mean atomic weight of the gas, $m_{H_2}$ is the mass of a hydrogen molecule, $d$ is the distance of the
source, $S_{\nu}$ is the flux density, and other symbols have common meanings. The mean excitation temperature
$T_{ex}$ is estimated to be 24 K from the NH$_3$(1,1), (2,2) emission. It should be noted that
$[\frac{H_{2}}{SiO}]$ has large uncertainty since the SiO abundance can be greatly enhanced by shocks as a result
of grain destruction leading to Si injection into the gas phase \citep{Seab83}. In contrast to the typical SiO
abundance $10^{-12}$ to $10^{-11}$ in dark clouds \citep{Ziurys89}, \citet{Mikami92} and \citet{Zhang95} detect
the SiO enhancement of four to five orders of magnitude toward the L1157 outflow. \citet{Hirano01} estimate the
SiO abundance to be $10^{-10}$ to $10^{-8}$ toward the multiple outflows in IRAS 16293-2422. We adopt an SiO
abundance of $10^{-8}$, which corresponds to an enhancement of three to four orders of magnitude with respect to
that in dark clouds. Then the outflow mass can be determined. Consequently we get the outflow rate according to
$$\dot{M}_{out}=\frac{M_{out}}{t_{dyn}}.$$ The derived results are listed in Table \ref{table1}.

\begin{figure*}
\plotone{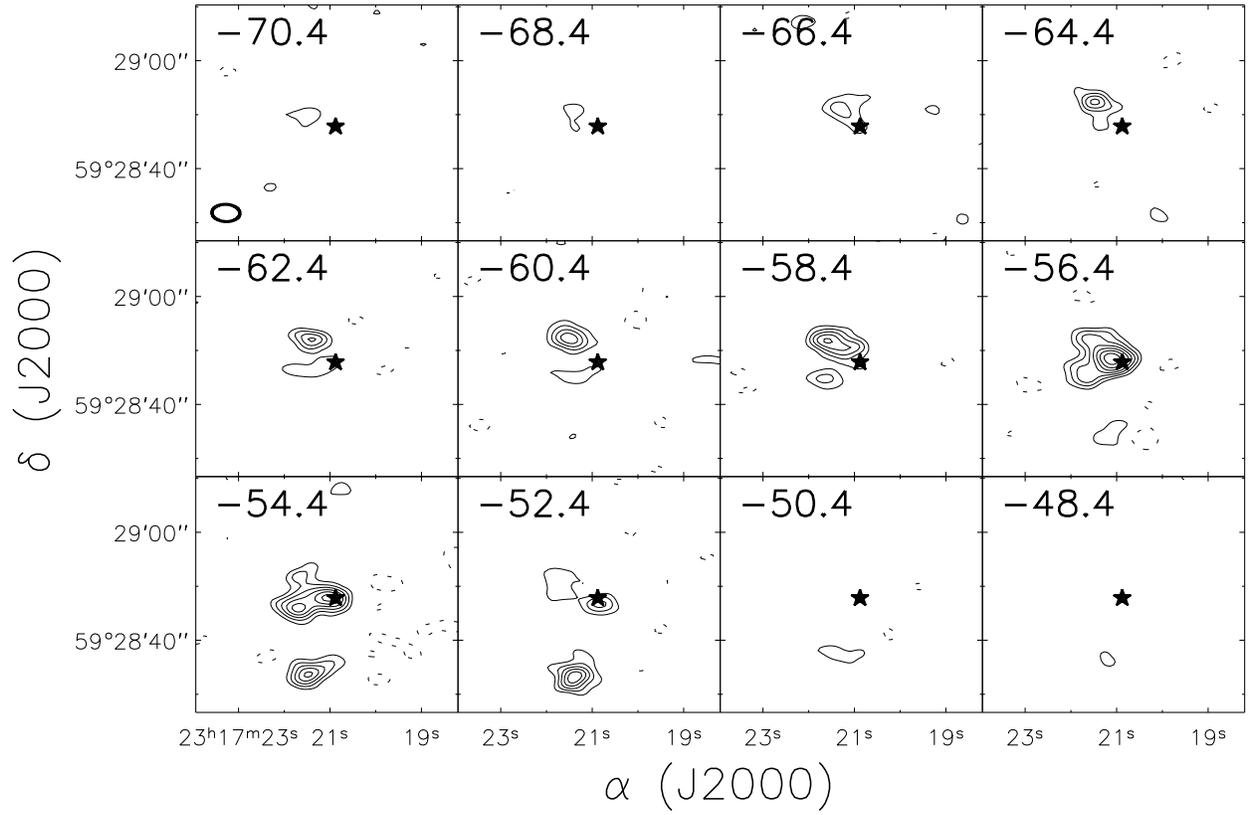} \caption{Same as in Fig. \ref{siochan1}, but for I23151 with 1 kms$^{-1}$ resolution. The
contours start at 16.5 mJy (3$\sigma$) in steps of 11 mJy (2$\sigma$). \label{siochan2}}
\end{figure*}

\begin{figure}
\plotone{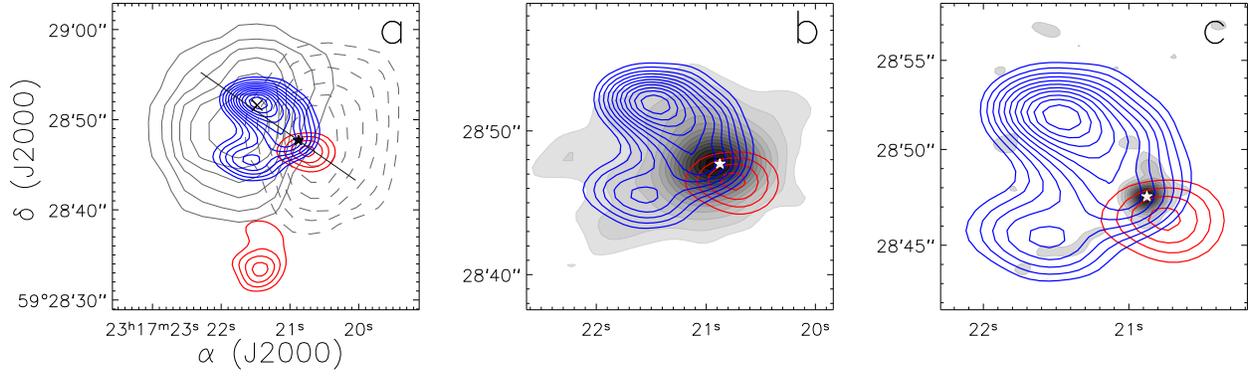} \caption{The blue and red contours represent the blue- and redshifted SiO emission in I23151,
which are integrated from -68.4 kms$^{-1}$ to -55.4 kms$^{-1}$ and from -53.4 kms$^{-1}$ to -50.4 kms$^{-1}$,
respectively. The blue contours star at 0.115 Jy$\cdot$kms$^{-1}$ (5$\sigma$) in steps of 0.046
Jy$\cdot$kms$^{-1}$ (2$\sigma$) and the red contours start at 0.08 Jy$\cdot$kms$^{-1}$ (5$\sigma$) in steps of
0.032 Jy$\cdot$kms$^{-1}$ (2$\sigma$). The solid and dashed grey contours in (a) describe the blue- and redshifted
CO outflow in single-dish observations \citep{Beuther02c}. The solid straight line and the cross mark the axis for
PV plotting and the site for the SiO spectrum drawing,respectively. The grayscales in (b) and (c) represent the
3.4 mm and 1.3 mm continuum emission, respectively. \label{sioint2}}
\end{figure}

The channel maps with 1 kms$^{-1}$ resolution in SiO in I23151 are presented in Fig. \ref{siochan2}. With respect
to the systemic velocity -54.4 kms$^{-1}$ ($v_{LSR}$), the SiO emission is mainly buleshifted and extends to the
velocity ${\Delta}v\sim-14$ kms$^{-1}$. The integrated emission is shown in Fig.\,\ref{sioint2}. In addition to
the emission close to the mm continuum source, there is a weaker redshifted feature to the south. This is maybe an
independent outflow from a low-mass (proto)star whose mass is below our detection limit. The bipolar SiO outflow
seems to emanate from the mm continuum peak and coincides with the respective inner parts of the CO outflow from
the single-dish observations (see Fig.\,\ref{sioint2}a). The dominant buleshifted SiO outflow has a
quasi-parabolic shape with the mm peak at its tip and a cavity in the center. This structure can also be
identified in the channel maps, especially in the -56.4 kms$^{-1}$ channel. Using the bispectrum speckle
interferometry method, \citet{Weigelt06} reveal a cone-like nebulosity in the near-infrared $K'$ band with a
bright near-infrared point source at the tip. This nebulosity coincides well with the quasi-parabolic SiO outflow
here. In Figs. \ref{sioint2}b and \ref{sioint2}c the quasi-parabolic SiO outflow also coincides with the
fan-shaped structure in the 3.4 mm continuum and the arch structure in the 1.3 mm continuum. The arch at 1.3 mm is
detected only at $2-3 {\sigma}$ level. However, the agreement with the 3.4 mm and the near-infrared image suggests
that the structure is real. Similar to I18264, we derive the characteristic parameters of the SiO outflow in
I23151, where $T_{ex}$ is assumed to be 30 K. The results are given in Table\,\ref{table1}.

\subsection{H$^{13}$CO$^+$ condensations} 
\begin{figure}
\plottwo{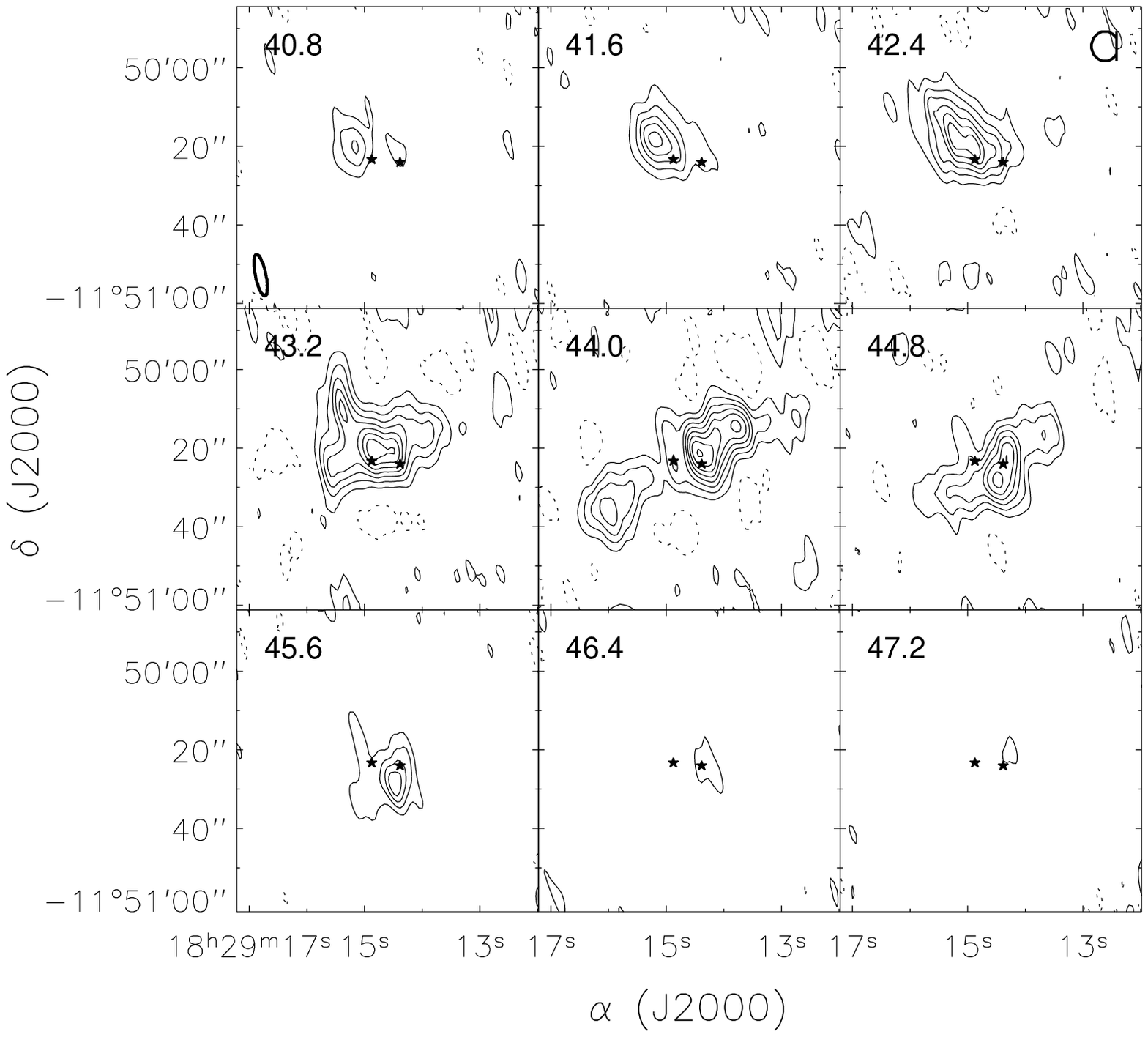}{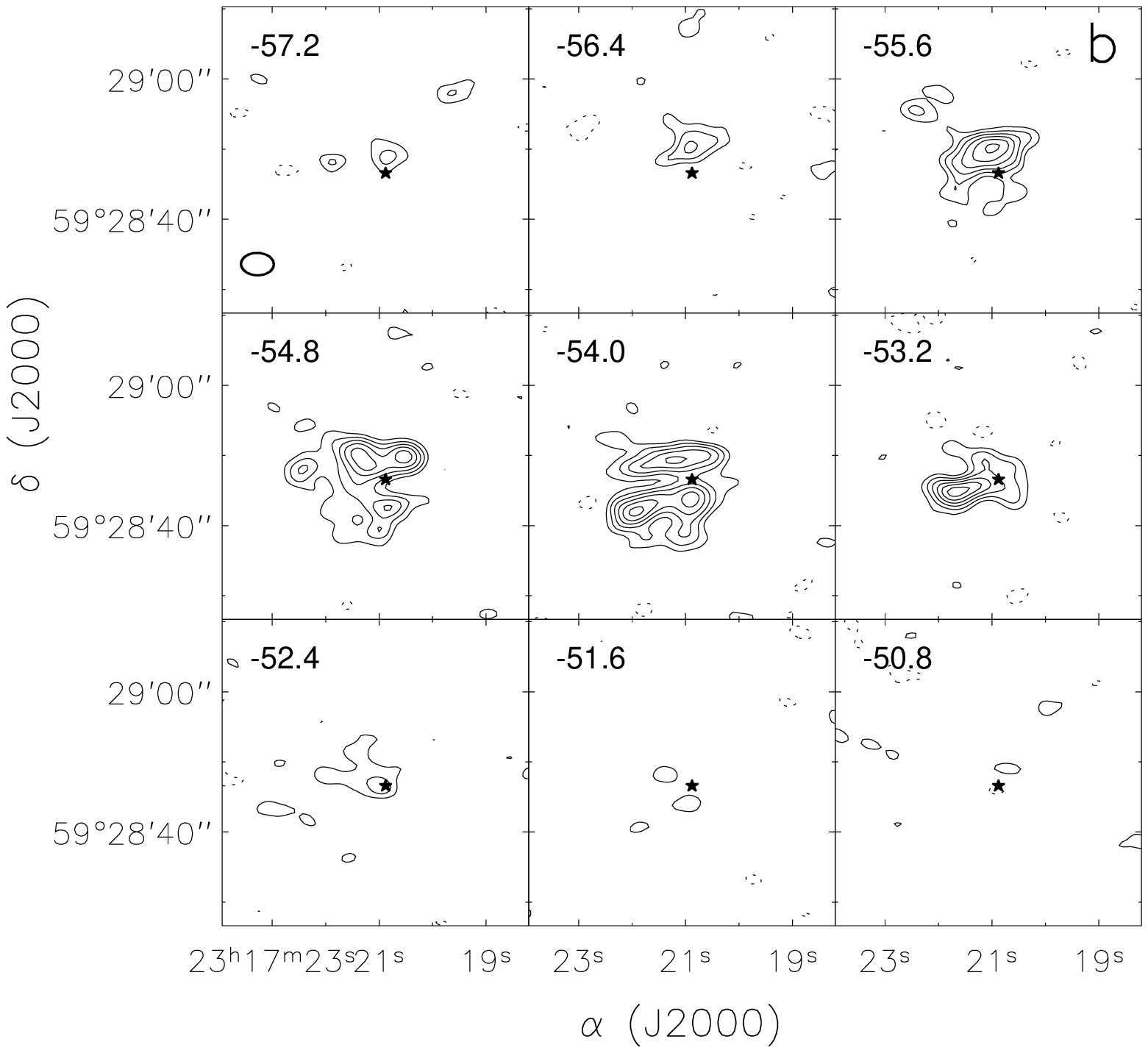} \caption{(a) The H$^{13}$CO$^{+}$ (1-0) channel maps for I18264. The
lowest and spacing contours are 78 mJy (3$\sigma$). The dashed contours represent the -3$\sigma$ level. The thick
ellipse in the lower left of the first panel shows the beam size. (b) Same as for (a), but for I23151. The
contours start at 21 mJy (3$\sigma$) in steps of 14 mJy (2$\sigma$). \label{hco}}
\end{figure}

Figs. \ref{hco}a and \ref{hco}b present the H$^{13}$CO$^+$ channel maps
with 0.4 kms$^{-1}$ resolution for I18264 and I23151, respectively. In both sources, the H$^{13}$CO$^+$ emission
is limited to velocity channels around the systemic velocity, which confirms that this line emission mostly traces
the dense ambient gas. For I18264, the peaks of the H$^{13}$CO$^+$ emission in the slightly blue- and redshifted
channels approximately coincide with the western and eastern peaks of the mm continuum, respectively. Figs.
\ref{hcoint}a and \ref{hcoint}b present the integrated emission overlaid with the SiO outflows for the two
sources. For I18264, the double peaks in the mm continuum are roughly resolved in the integrated H$^{13}$CO$^+$
emission. In addition to the condensations around the mm continuum peaks, the H$^{13}$CO$^+$ emission also shows
extensions coincident with the SiO outflows. In particular the southeast elongated feature in H$^{13}$CO$^+$
coincides well with the jet-like SE SiO outflow. For I23151, the major and minor peaks in the integrated
H$^{13}$CO$^+$ emission correlates well with the SiO clumps. The quasi-parabolic shaped SiO outflow also has
counterparts in H$^{13}$CO$^+$, especially in the -54.0 kms$^{-1}$ channel and integrated emission. The
correlation between the integrated H$^{13}$CO$^+$ emission and SiO emission for the two sources suggests that the
H$^{13}$CO$^+$ gas may be influenced by the outflows. Adopting an HCO${^+}$ abundance of $1\times10^{-9}$
\citep{Dishoeck93} and a C to $^{13}$C ratio of 67 \citep{Langer90}, we can estimate the gas masses of the
H$^{13}$CO$^+$ condensations with the similar method used in the estimation of the gas masses in the SiO outflows.
The results ($M_{dense}$) are listed in Table \ref{table1}. Note that the masses derived from the H$^{13}$CO$^+$
emission are much larger than the core masses derived from the mm continuum. The integrated H$^{13}$CO$^+$
emission is much more extended than the interferometric mm continuum for both sources. Based on the single-dish
1.2 mm continumm observations, \citet{Beuther02b} derive the masses ${\sim}2100M_{\odot}$ and
${\sim}28000M_{\odot}$ respectively for the near and far distances for I18264 and ${\sim}620M_{\odot}$ for I23151,
which are consistent with the the masses of the H$^{13}$CO$^+$ condensations here. Thus the interferometric mm
dust continuum probably traces the densest cores, filtering out most emission and hence yielding less masses,
whereas the H$^{13}$CO$^+$ traces lower density gas and may even be slightly optically thick. In addition, the
H$^{13}$CO$^+$ emission is obviously affected by the outflows and HCO$^+$ and H$^{13}$CO$^+$ can be enhanced by
shocks. Toward the bipolar molecular outflow of L1157, \citet{Bachiller97} detect the abundance of HCO$^+$ to be
enhanced by a factor of $26 - 30$. \citet{Jorgensen04} reveal an HCO$^+$ abundance of $2.9{\times}10^{-9}$ in the
NGC 1333 outflow region. \citet{Girart05} derive an averaged HCO$^+$ abundance of $3.2{\times}10^{-9}$ over the
core located ahead of HH2. Thus the abundance of HCO$^+$ can be enhanced by a factor of ${\sim}\,3-30$. Assuming
the H$^{13}$CO$^+$ to HCO$^+$ ratio to be constant, the abundance of H$^{13}$CO$^+$ may be underestimated, and
consequently the mass is overestimated by a factor of ${\sim}\,3-30$.

\subsection{Ammonia emission} \label{ammonia} 
\begin{figure}
\plottwo{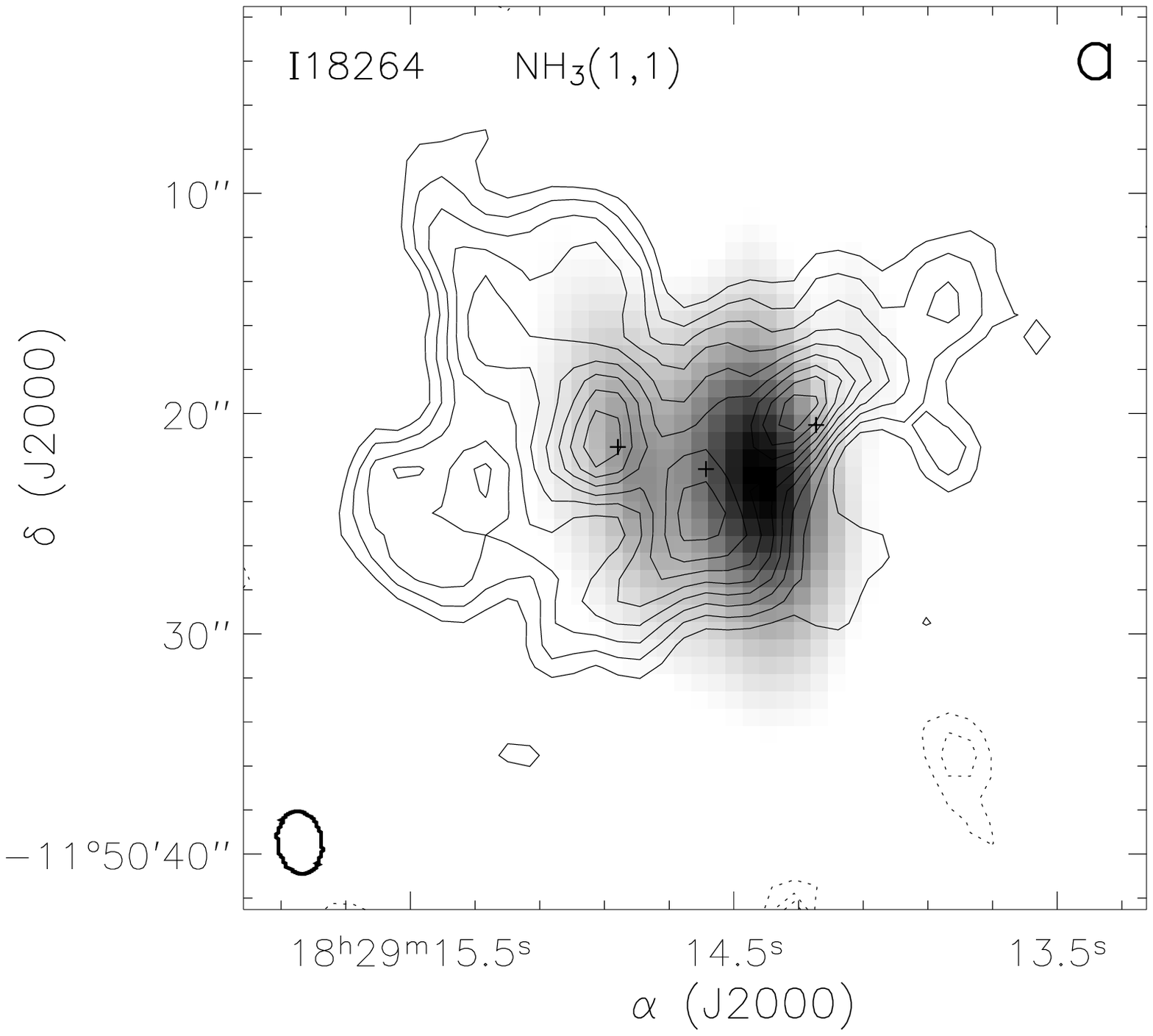}{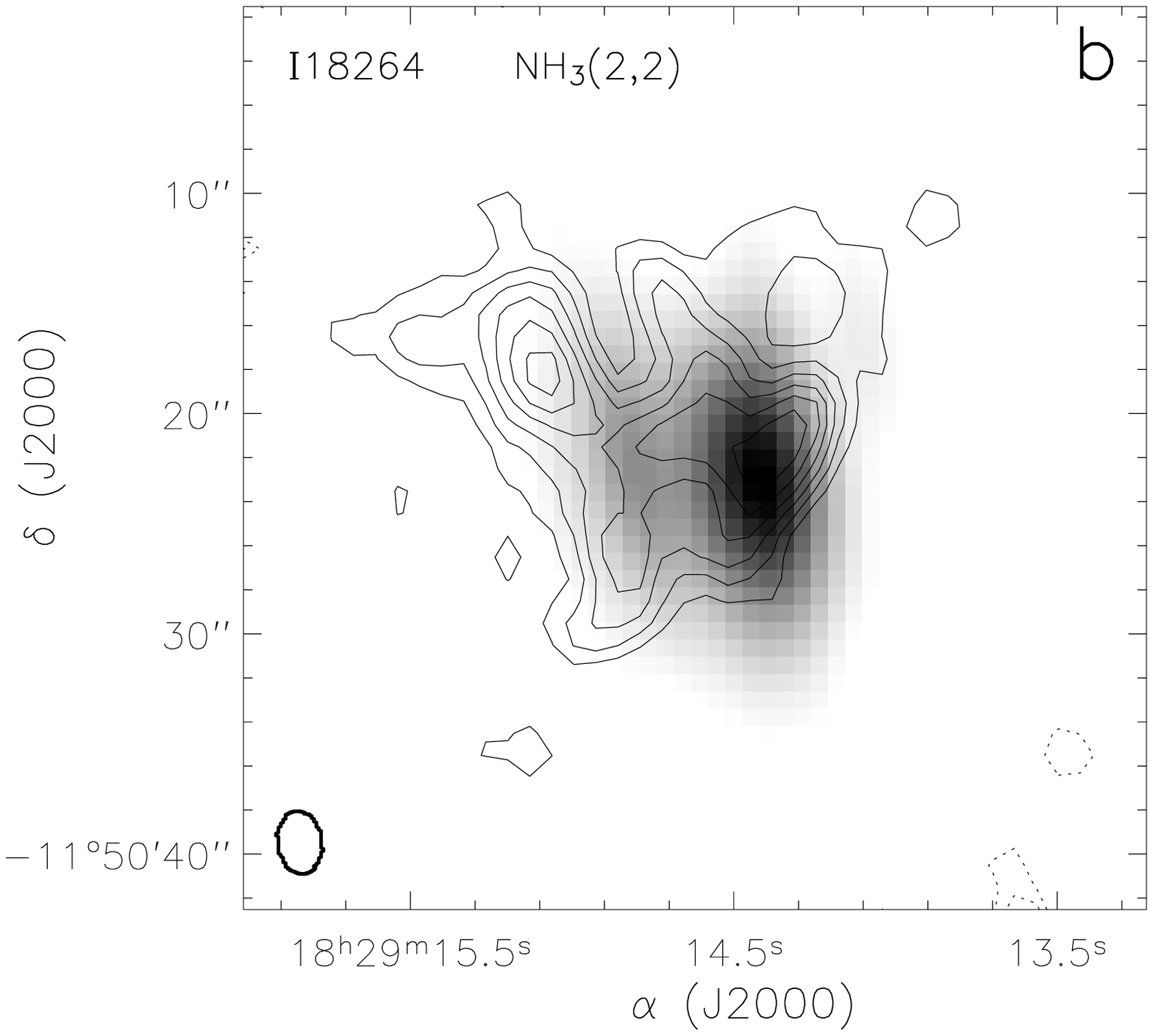} \caption{(a) The NH$_3$(1,1) emission in I18264 integrated from 41.8 kms$^{-1}$ to 45.4
kms$^{-1}$. The contours start at 20 mJy$\cdot$kms$^{-1}$ in steps of 13.4 mJy$\cdot$kms$^{-1}$. The pluses denote
the positions at which the physical parameters in Table. \ref{table2} are calculated. (b) Same as for (a), but for
NH$_3$(2,2) emission integrated from 41.6 kms$^{-1}$ to 44.6 kms$^{-1}$. The levels of contours are from 19.5
mJy$\cdot$kms$^{-1}$ in steps of 13 mJy$\cdot$kms$^{-1}$. The dotted contours represent the -3$\sigma$ and
-6$\sigma$ levels. The grayscale represent the 3.4 mm continuum. The beam sizes are marked at the lower left of
each panel. \label{nh3}}
\end{figure}

\citet{Sridharan02} detected the
NH$_3$(J,K)=(1,1), (2,2) inversion lines toward I18264 with the Effelsberg 100m telescope. The NH$_3$ emission
toward I23151 was also detected with the Effelsberg telescope, but is about an order of magnitude weaker than for
I18264 (Beuther, priv com.). With the VLA we observed the NH$_3$(1,1) and (2,2) lines toward the two regions and
detected the NH$_3$(1,1) and (2,2) emission only in I18264. We derive the optical depth $\tau(1,1,m)$ for the main
component of the NH$_3$(1,1) line and the rotational temperature $T_{rot}$(2,2:1,1) \citep{Ho83} at three
positions as denoted in Fig. \ref{nh3}a. The kinetic temperature $T_{kin}$ can be estimated according to
$$T_{rot}(2,2:1,1)=\frac{T_{kin}}{1+(T_{kin}/41.7)ln[1+C(2,2\rightarrow2,1)/C(2,2\rightarrow1,1)]},$$
where $C(2,2\rightarrow2,1)$ and $C(2,2\rightarrow1,1)$ are the rates of collisional transition between the levels
(2,2),(2,1) and (1,1) \citep{Walmsley83, Danby88}. The excitation temperature $T_{ex}$ is determined following
$$T_B(1,1,m)=(T_{ex}-2.7)(1-e^{-\tau(1,1,m)}),$$ where $T_B(1,1,m)$ is the brightness temperature of the main
component of the NH$_3$(1,1) line. Then the gas density $n(H_2)$ can be calculated according to
$$n(H_2)=\frac{A}{C}\left[\frac{T_{ex}-2.7}{T_{kin}-T_{ex}}\right]\left[1+\frac{kT_{kin}}{h\nu}\right],$$ where $A$
and $C$ are the Einstein $A$ coefficient and the collision rate, respectively \citep{Ho83}. The results derived
from the ammonia emission are listed in Table \ref{table2}. The derived mean rotation temperature 26 K is a little
higher than that of 18 K derived by \citet{Sridharan02}, and the mean kinetic temperature 36 K is approximately
equal to the dust temperature \citep{Sridharan02}. In Fig. \ref{nh3}, there are peaks in NH$_3$(1,1) and (2,2)
approximately coincident the mm continuum peaks. It seems that NH$_3$(1,1) and (2,2) lines trace a more extended
envelope than the mm continuum. The derived gas density is $\sim1.4\pm0.6\times10^5$ cm$^{-3}$ and the intrinsic
line width (FWHM) is ${\sim}2.5$ kms$^{-1}$.

\section{Discussion} \label{dis}
\subsection{Outflow Energetics}
In I18264, the SiO emission reveals two quasi-perpendicular molecular outflows emanating from the western mm peak.
The mass estimated from the mm continuum is as high as 570 $M_{\odot}$ and 7300 $M_{\odot}$ respectively for the
near and far distances even with a fraction of the flux being filtered out by the interferometer. Since the
western peak is much stronger than the eastern one, most of the mass is attributed to the western peak. To check
the missing short spacings, the convolved PdBI SiO data have been compared with the single-dish SiO data observed
with the IRAM 30m telescope (Beuther, priv com.). We find that at channels around the systemic velocity 43.6
kms$^{-1}$, less than 34\% of the SiO flux is filtered out by the interferometer. While at channels above
$v_{LSR}$ of 50 kms$^{-1}$, no missing short spacing problem occurs. The outflow mass for I18264 in Table
\ref{table1} is derived from channels of $33.6-40.6$ kms$^{-1}$ ($M_{blue}$) and $46.6-103.6$ kms$^{-1}$
($M_{red}$), which are both 3 kms$^{-1}$ apart from the systemic velocity. So the derived outflow mass for I18264
does not suffer much from missing short spacings, and the results are comparable to that derived from the
single-dish CO observations \citep{Beuther02c}. From the SiO channel maps and the integrated emission $M_{red}$ is
mostly due to the jet-like SE outflow, and gives mass outflow rates of $3.4{\times}10^{-3}M_{\odot}\,yr^{-1}$ and
$1.2{\times}10^{-2}M_{\odot}\,yr^{-1}$ for the near and far distances, respectively. With the assumption of the
momentum conservation between the observed outflow and the internal jet, and adopting a typical jet velocity of
500 kms$^{-1}$, the mass loss rates caused by the underlying jet or wind are respectively about
$4.1\times10^{-4}M_{\odot}\,yr^{-1}$ and $1.4\times10^{-3}M_{\odot}\,yr^{-1}$. Assuming further a ratio between
the mass loss rate and accretion rate of approximate 1/3 \citep{Shu87, Tomisaka98}, we get accretion rates of
$1.2\times10^{-3}M_{\odot}\,yr^{-1}$ and $4.2\times10^{-2}M_{\odot}\,yr^{-1}$ respectively for the near and far
distances. Such accretion rates are high enough to overcome the radiation pressure of the central (proto)star and
form most massive stars \citep{Wolfire87, Jijina96, Yorke02}. For I23151, the estimation from the mm continuum
gives a core of 170 $M_{\odot}$. With the same assumptions as for I18264, the outflow rate
$2.3\times10^{-4}M_{\odot}\,yr^{-1}$ will lead to an accretion rate of $1.9\times10^{-5}M_{\odot}\,yr^{-1}$.
According to the comparison between the convolved PdBI SiO data and that from the IRAM 30m single-dish
observations (Beuther, priv com.), about $32\%-43\%$ of the flux is filtered out by the interferometer for most
velocity channels. Considering the missing flux due to the missing short spacings and the outflow mostly being
blueshifted, the accretion rate can be underestimated. Then the accretion rate for I23151 can also be high enough
to overcome the radiation pressure and form massive stars \citep{Wolfire87, Jijina96, Yorke02}.

\subsection{Mass-velocity diagrams} \label{mvd}
\begin{figure}
\plottwo{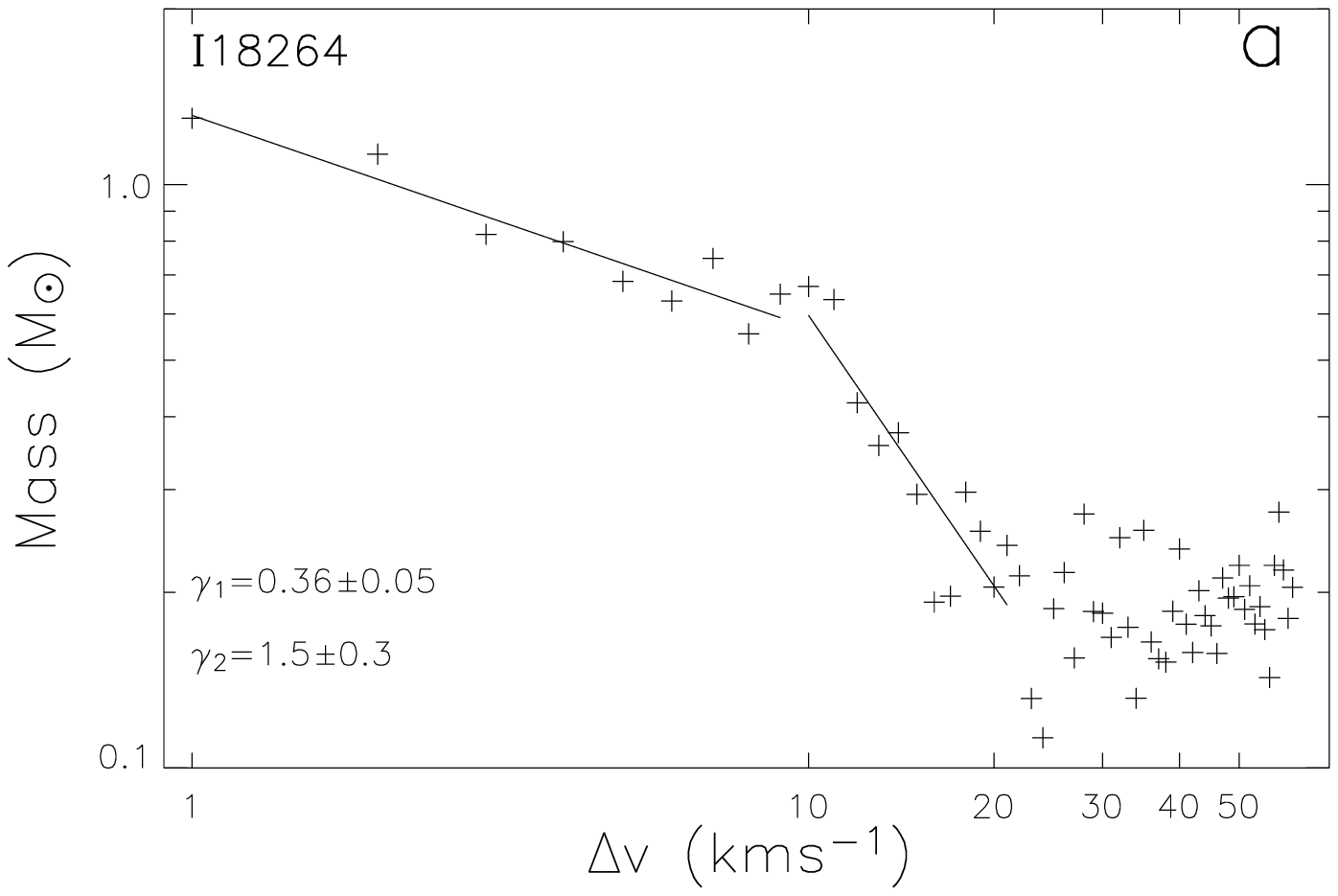}{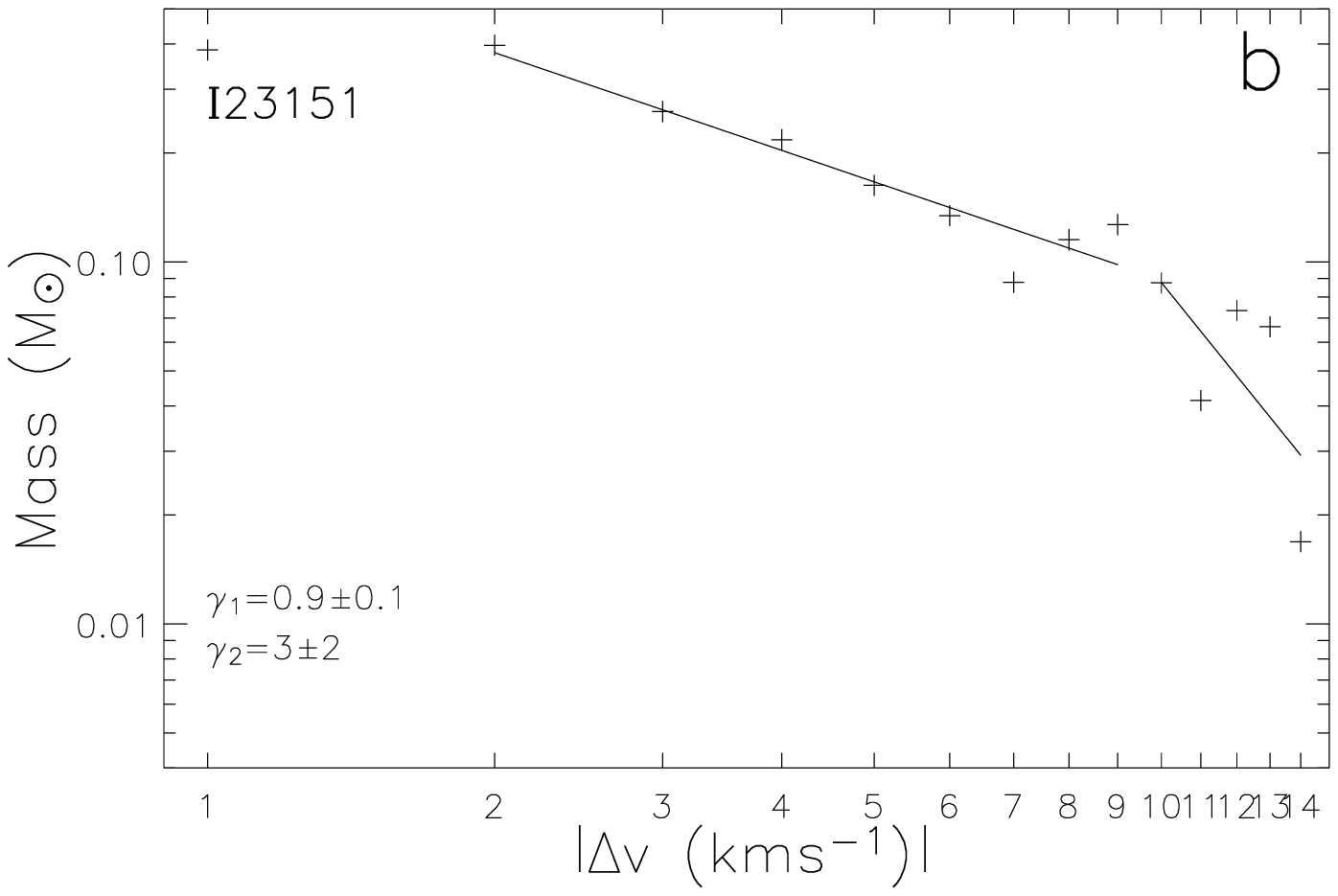} \caption{(a)Outflow masses calculated for the near distance for 1
kms$^{-1}$ channels as a function of flow velocities for I18264. Note outflow masses for the far distance will
give the same broken power law. (b) Same as for (a), but for I23151. The solid lines show the power law fits of
$m(v) \propto v^{-\gamma}$. \label{mv}}
\end{figure}

Outflows associated with low-mass young stellar objects (YSOs) usually exhibit a ``mass spectrum'' $m_{CO}(v)
\propto v^{-\gamma}$ \citep[e.g.][]{Chandler96, Lada96}. The power law index $\gamma$ is typically $\sim1.7-1.8$
\citep{Lada96}, although the slope often steepens at velocities greater than 10 kms$^{-1}$ from the systemic
velocity. From a compilation of 22 sources with luminosities ranging from 0.58 $L_{\odot}$ to
$3\times10^5\,L_{\odot}$, \citet{Richer00} find that at velocities below 10 kms$^{-1}$ $\gamma$ is similar, while
at velocities above 10 kms$^{-1}$, $\gamma$ is $3-4$ for low-mass YSOs and $3-8$ for luminous YSOs. Toward a
sample of 11 objects with luminosities $L_{bol}\,{\geq}\,10^2L_{\odot}$ at a distance of 2 kpc, \citet{Ridge01} do
not find a correlation between mass-spectrum slope and bolometric luminosity nor a clear separation between
$\gamma$s measured for the high velocity and low velocity emission. \citet{Su04} also study the mass-velocity
diagrams of outflows of two luminous YSOs and find a change in slope at a break-point velocity of 10 kms$^{-1}$.
They suggest that the high velocity ($|{\Delta}v|>10$ kms$^{-1}$) gas may drive the low velocity ($|{\Delta}v|<10$
kms$^{-1}$) gas. With the assumption of the optically thin SiO thermal emission in LTE, we derive the
mass-velocity diagrams for the SiO outflows for the two sources (see Fig. \ref{mv}). Since the SiO emission is
dominantly redshifted in I18264 and blueshifted in I23151, we derive the mass-velocity relations in the
corresponding wings for the two sources. The redshifted emission in I18264 is mostly attributed to the SE outflow.
So Fig. \ref{mv}a represents the mass-velocity relationship of the HC of the SE outflow and can be fitted by a
broken power law with the index steepening from ${\gamma}_1=0.36\,{\pm}\,0.05$ to ${\gamma}_2=1.5\,{\pm}\,0.3$ at
${\Delta}v=10$ kms$^{-1}$. The mass distribution in the velocity channels of the VHC of the SE outflow has large
scatters and can hardly be described by a linear fit. Similarly, the blueshifted lobe of the SiO outflow in I23151
is also fitted by a broken power law with $\gamma_1=0.9\,{\pm}\,0.1$ for $|{\Delta}v|<10$ kms$^{-1}$ and
$\gamma_2=3\,{\pm}\,2$ for $|{\Delta}v|{\geq}10$ kms$^{-1}$. From three dimensional simulations of a dense
molecular jet penetrating a dense molecular medium, \citet{Smith97} predict the change in slope at high velocities
due to a jet-bow shear layer consisting of molecules which survived the jet terminal shock. Models of material
being accelerated by jet-driven bow shocks \citep{Downes99} have reproduced a power law relationship between mass
and velocity with ${\gamma}$ increasing with decreasing molecular abundance in the jet. However, \citet{Downes99}
give an upper limit of 3.75 for ${\gamma}$, which is much less than the observed values. Therefore the physical
origin is still unclear for the broken power law of the mass-velocity diagrams.

\subsection{Kinematics} \label{kin}
\begin{figure}
\plotone{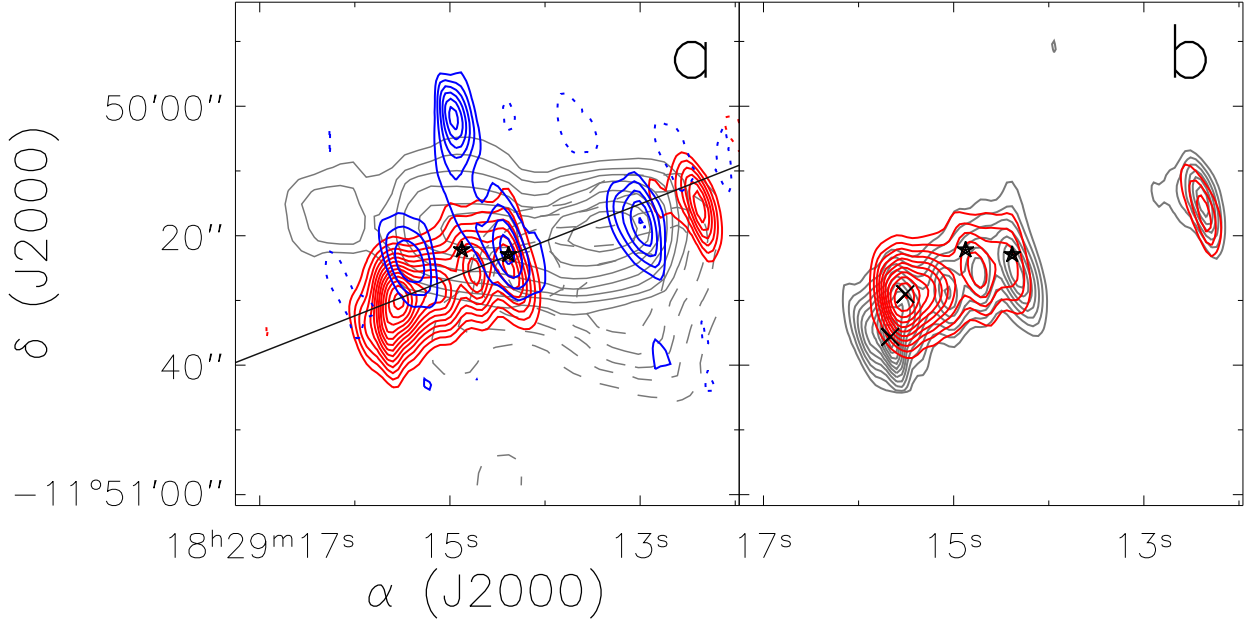} \caption{The integrated SiO emission in I18264. (a) The blue and red contours show the blue- and
redshifted SiO emission, respectively. The blueshifted emission is integrated from 33.6 kms$^{-1}$ to 40.6
kms$^{-1}$ with contour levels starting at 0.21 Jy$\cdot$kms$^{-1}$ (3$\sigma$) in steps of 0.14
Jy$\cdot$kms$^{-1}$ (2$\sigma$). The redshifted emission is integrated from 46.6 kms$^{-1}$ to 104.6 kms$^{-1}$
with contour levels starting at 0.6 Jy$\cdot$kms$^{-1}$ (3$\sigma$) in steps of 0.4 Jy$\cdot$kms$^{-1}$
(2$\sigma$). The blue and red dashed contours represent the -3$\sigma$ level of the blue- and redshifted emission,
respectively. The solid and dashed grey contours represent the blue- and redshifted lobes of the CO outflow in
single-dish observations \citep{Beuther02c}. The solid straight line denotes the axis along which the PV diagram
is plotted. (b) The gray and red contours denote the redshifted emission integrated from the HC channels and from
the VHC channels, respectively. The gray contours start at 0.339 Jy$\cdot$kms$^{-1}$ (3$\sigma$) in steps of 0.226
Jy$\cdot$kms$^{-1}$ (2$\sigma$), and the red contours start at 0.486 Jy$\cdot$kms$^{-1}$ (3$\sigma$) in steps of
0.324 Jy$\cdot$kms$^{-1}$ (2$\sigma$). The two crosses mark the sites at which the SiO spectra are drawn.
\label{sioint1}}
\end{figure}

\begin{figure}
\plotone{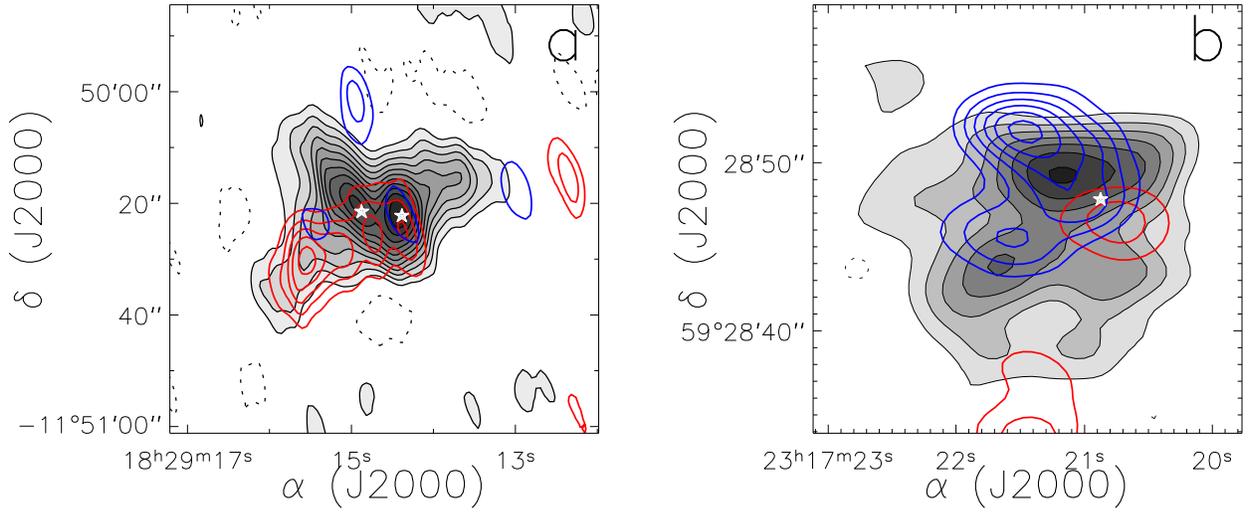} \caption{(a) The grayscale and contours represent the H$^{13}$CO$^{+}$ (1-0) emission integrated
from 41.2 to 45.4 kms$^{-1}$ for I18264, starting at 192 mJy$\cdot$kms$^{-1}$ (4$\sigma$) in steps of 144
mJy$\cdot$kms$^{-1}$ (3$\sigma$). The dashed contours represent the -4$\sigma$ level. The blue and red contours
show the SiO outflow same as in Fig. \ref{sioint1}a  but at 5$\sigma$, 10$\sigma$, 15$\sigma$, 20$\sigma$,
25$\sigma$ levels. (b) Same as for (a), but for I23151. The grayscale and contours represent the H$^{13}$CO$^{+}$
(1-0) emission integrated from -56.0 to -52.8 kms$^{-1}$, starting at 36 mJy$\cdot$kms$^{-1}$ (3$\sigma$) in steps
of 36 mJy$\cdot$kms$^{-1}$. The blue and red contours represent the integrated SiO emission same as in Fig.
\ref{sioint2} but at 5$\sigma$, 9$\sigma$, 13$\sigma$, 17$\sigma$, 21$\sigma$, 25$\sigma$ levels. \label{hcoint}}
\end{figure}

\begin{figure}
\plotone{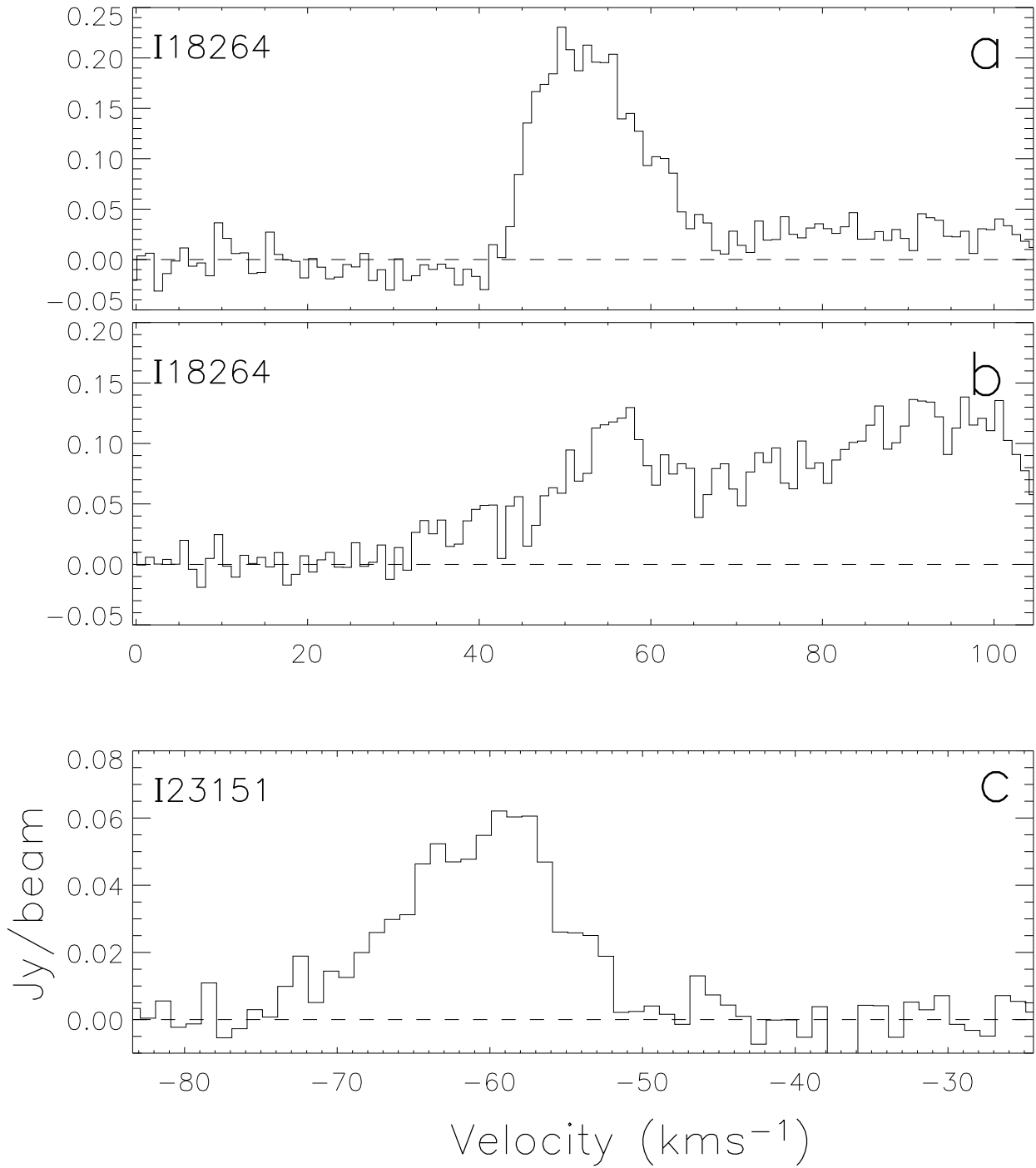} \caption{The SiO line profiles. (a)The SiO line profile of I18264 at the position denoted as the
southern cross in Fig. \ref{sioint1}b; (b)Same as for (a), but at the position denoted as the northern cross in
Fig. \ref{sioint1}b; (c)The SiO line profile of I23151 at the position denoted as a cross in Fig. \ref{sioint2}a.
\label{profile}}
\end{figure}

The SiO in the gas phase of molecular outflows can be produced through the sputtering of Si-bearing material in
grains, where the sputtering is driven by neutral particle impact on charged grains in shocks \citep{Schike97}. In
Fig. \ref{hcoint}a, the jet-like SE outflow in I18264 coincides well with the H$^{13}$CO$^+$ structure extending
to the southeast, suggesting the SiO emission in this region may arise from the interaction between the shocks
driven by the jet or wind from the central (proto)star and the dense ambient gas clump. Such an interaction mainly
exists in the redshifted lobe since the SiO emission is dominantly redshifted. In Fig. \ref{profile}a, the SiO
line at the downstream peak of the HC of the SE outflow (denoted as the southern cross in Fig. \ref{sioint1}b)
shows a profile with a steep decrease toward the systemic velocity and a gradual redshifted wing, which suggests
the SiO enhancement arising from the quiescent material accelerated by the shock when the jet or wind impinges
into the dense ambient gas. We have checked the corresponding single-dish SiO spectrum and find that such a
profile also exists. So the steep decrease toward the systemic velocity can not be due to the missing short
spacings. This kind of SiO profiles and interactions have been observed toward several low-mass outflows
\citep[e.g.][]{Zhang95,Hirano01}. The line profile at the downstream peak of the VHC (denoted as the northern
cross in Fig. \ref{sioint1}b) shows a very broad plateau in the redshifted wing with velocities up to about 60
kms$^{-1}$ from the systemic velocity (see Fig. \ref{profile}b). Such a broad plateau is consistent with the
flattened mass spectrum of the VHC in the mass-velocity diagram. For I23151, there is a good correlation between
the SiO clumps and the integrated H$^{13}$CO$^+$ peaks. The SiO line at the peak of the blueshifted lobe (denoted
as a cross in Fig. \ref{sioint2}) also shows a profile of a steep drop toward the systemic velocity and a gradual
decline in the blueshifted wing (see Fig. \ref{profile}c), and this profile is confirmed in the single-dish data.
So the SiO emission in I23151 can also arise from the interaction between the jet or wind driven by the central
(proto)star and the dense ambient gas.

Toward several low luminosity sources (e.g., L1448, \citet{Guilloteau92}; NGC 2264G, \citet{Lada96}), it is
observed that the flow velocity increases with distance from the driving source. On the contrary, \citet{Zhang00}
find the SiO outflow velocities in L1157 decreasing toward the clumps farther away from the driving source. Also
considering the different orientations of the different pairs of SiO clumps, they explain this in terms of
multiple shocks driven by a processing and episodic jet. The downstream peak of the HC of the SE outflow in I18264
is farther away from the driving source than that of the VHC, which is similar to the SiO outflow in L1157. And
the VHC of this outflow is northward offset from the HC. We suggest that the SE outflow is driven by a precessing
jet. In this scenario, the lower velocities of the HC as compared with the VHC imply a deceleration of the
underlying jet after travelling a larger distance; The VHC is younger and driven by newly formed shocks when the
underlying jet is precessing to the north, and its flattened mass spectrum and the very broad line width suggest
it does not experience significant deceleration. In addition, the VHC of the northwest SiO clump is also slightly
shifted to the south of the HC, and both the VHC and HC align in line with the VHC and HC of the SE outflow,
respectively. This also suggests the precessing scenario. Most recently, the VLA observations by \citet{Zapata06}
resolve the western mm peak in I18264 into triple sources at 1.3 cm and 7 mm. One of the three sources shows a
slightly rising spectral index between 3.6 cm and 1.3 cm and a flux density at 7 mm larger than the value
extrapolated from the 3.6 cm and 1.3 cm observations. \citet{Zapata06} suggest this source to be a combination of
a thermal jet and disk. This is consistent with our interpretation that the SE outflow being driven by a
precessing jet.

In case of I23151, the SiO outflow in I23151 traces the inner parts of the large bipolar CO outflow despite the
additional feature to the south. As described above, the blueshifted outflow is revealed as a quasi-parabolic
structure, which is coincident with the similar structures in the 3.4 mm continuum and H$^{13}$CO$^+$ line
emission and also coincident with the near-infrared cone-like nebulosity by \citet{Weigelt06}. Apparently, this
can be interpreted as the molecular outflow being entrained by a wide-angle wind, where the low-intensity SiO
emission in the central region of the cone is due to the cavity cleared by the wind while the enhanced SiO
emission at the base of the outflow is caused by the interaction between the wind and the ambient material in the
wall of the cavity. \citet{Weigelt06} also reveal a bright point source at the tip of the cone-like nebulosity. As
the point source locating at the center of the single-dish 1.2 mm peak, \citet{Weigelt06} suggest it to be the
protostar driving an outflow traced by the nebulosity. The astrometry of the point source is (RA,
DEC)(J2000)=(23:17:21.02, 59:28:48.0) from the comparison of the speckle image with the 2MASS images and 2MASS
All-Sky Catalog of Point Sources. The position of the mm peak of our PdBI observations is (RA,
DEC)(J2000)=(23:17:20.88, 59:28:47.7). Considering the 2MASS positional accuracy of $1''$, the PdBI mm peak can be
the same as the near-infrared point source. But it is also possible that the offset is real and the very bright
near-infrared point source is too evolved to be a protostar and the driving source of the outflow corresponds to
the interferometric mm peak which is not resolved in the single-dish 1.2 mm continuum with an $11''$ resolution.

\begin{figure}
\plottwo{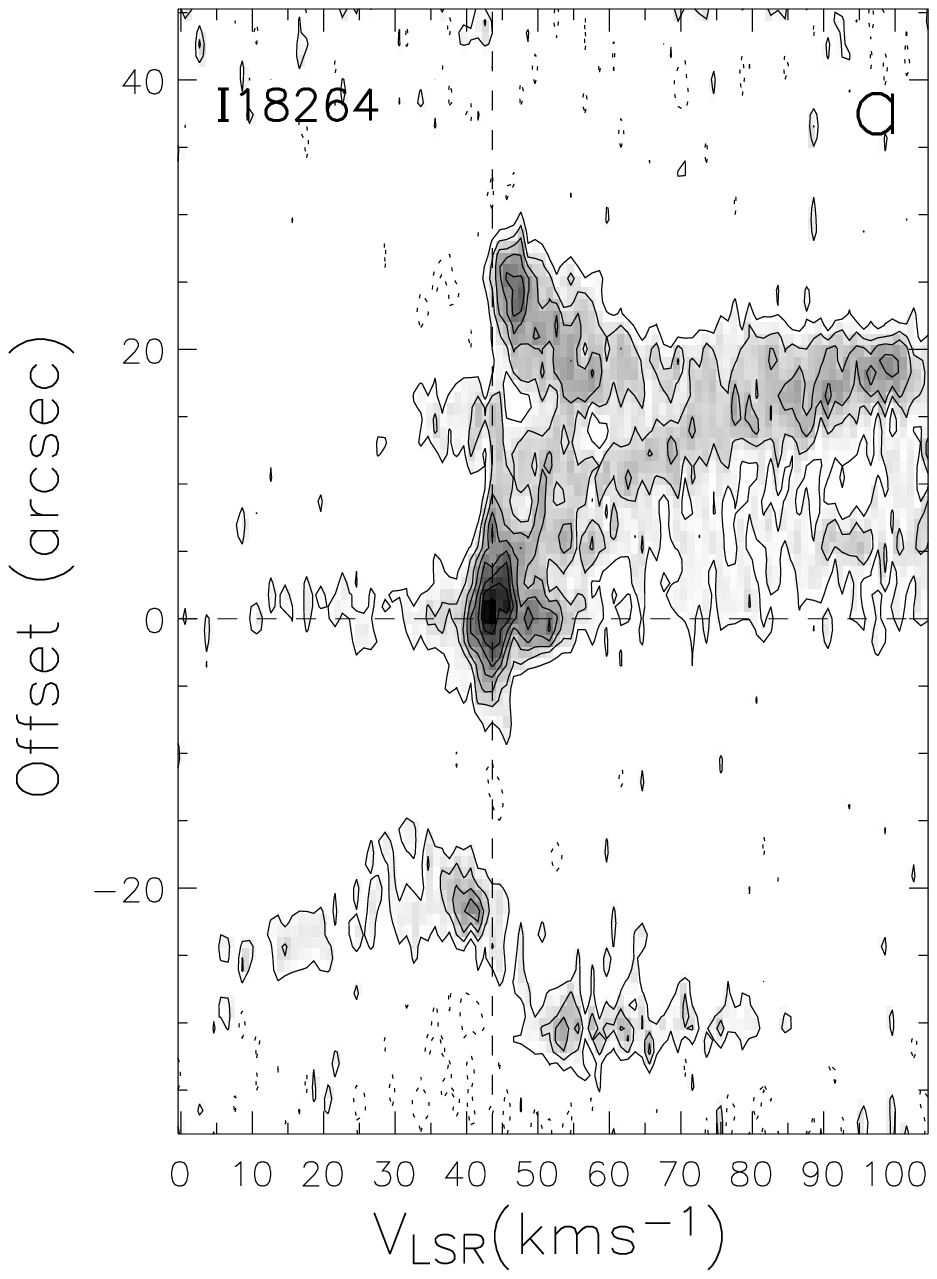}{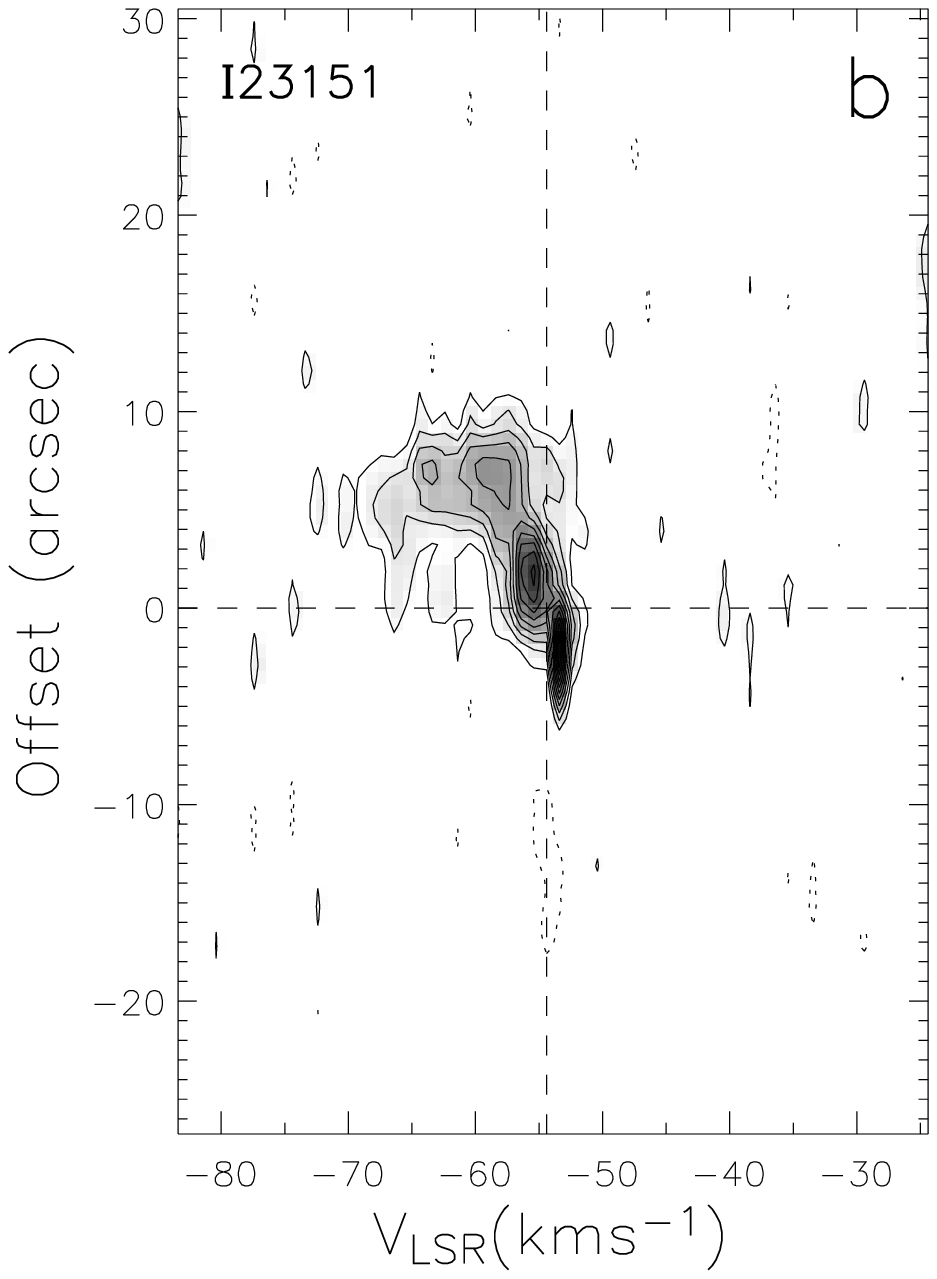} \caption{(a) The SiO position-velocity diagram for I18264 along the axis denoted as a
straight line in Fig. \ref{sioint1}a. The greyscale and solid contours start at $2\sigma$ in steps of $2\sigma$.
The dashed contours mark the -2$\sigma$ level. (b) Same as (a), but for I23151 and along the axis denoted as a
straight line in Fig. \ref{sioint2}a. The contour leves start at 2.5$\sigma$ in steps of $2\sigma$ and the dashed
contours represent the $-2.5\sigma$ level. \label{pv}}
\end{figure}

The position-velocity (PV) diagrams of the SiO outflows along the axes marked in Fig. \ref{sioint1}a and Fig.
\ref{sioint2}a are shown in Fig. \ref{pv}. For I18264, in addition to the feature around the driving source with
velocities near the systemic velocity, there are mainly two components in the PV diagram: the component with
velocities increasing linearly with the distance from the driving source; the component farther away from the
driving source with maximum velocities decreasing with the increasing distance. Compared with the channel maps in
Fig. \ref{siochan1}, we can find that these two components correspond to the VHC and HC of the SE outflow in Fig.
\ref{sioint1}b. There is a pair of faint feature centered at about $-27''$, which corresponds to the NW lobe of
the SE-NW outflow. The velocity structure of the peak emission of the VHC follows the so called ``Hubble law'' and
can be explained in terms of bow-shock acceleration. This is compatible with the interpretation that the VHC is
excited by the newly formed shocks without undergoing significant deceleration. If the broken power law is caused
when the low velocity ($|{\Delta}v|<10$ kms$^{-1}$) gas comprises ambient gas entrained by the high velocity
($|{\Delta}v|>10$ kms$^{-1}$) gas \citep{Su04}, the high velocity gas should inject momentum into the low velocity
gas and decelerate when it travels downstream. This scenario is in agreement with the velocity structure of the HC
in the PV diagram of I18264, where the maximum velocities decrease with the distance from the driving source. For
I23151, the PV diagram roughly follows the Hubble law. At the position about $7''$ from the driving source,
emission features have a relatively wide range of velocities.

\section{Conclusions} \label{conclusion}
We carry out a study of SiO outflows toward the high-mass star forming regions I18264 and I23151. According to the
H$^{13}$CO$^+$ observations and SiO line profiles, the SiO emission in these two sources can be caused by the
interaction between the jet or wind from the central young stars and the dense ambient gas. The mass-velocity
relation of the HC of the SE outflow in I18264 and the blueshifted outflows in I23151 also can be fitted by a
broken power law with the slopes steeping at about 10 kms$^{-1}$ with respective to the systemic velocity. With
the PV characteristic of the terminal velocities decreasing with the distance from the driving source, the low
velocity ($|{\Delta}v|<10$ kms$^{-1}$) gas probably comprises the ambient gas entrained by the high velocity
($|{\Delta}v|>10$ kms$^{-1}$) gas.

Toward I18264, at least two quasi-perpendicular outflows with high collimation factors ($\sim3-4$) are resolved.
The VHC of the SE outflow has very high velocities up to ${\Delta}v\sim60$ kms$^{-1}$. It is north offset from the
HC, and its downstream peak is farther away from the driving source. Comparing with the outflows in the well
studied low-mass source L1157, we suggest the SE outflow can be entrained by an underlying precessing jet. The
characteristics of the HC and VHC of the SE outflow in the mass-velocity and PV diagrams also support the
processing jet scenario. For I23151, the blueshifted SiO emission traces a quasi-parabolic shaped outflow which
can be identified in the 3.4 mm continuum and H$^{13}$CO$^+$ emission and coincide with the near-infrared
nebulosity in the literature. This outflow can be interpreted as the molecular gas entrained by the underlying
wide-angle wind.

The core masses estimated from the 3.4 mm continuum are 570 $M_{\odot}$ and 7300 $M_{\odot}$ respectively for the
near and far distances for I18264 and 170 $M_{\odot}$ for I23151. With the assumption of the momentum conservation
between the outflow and the driving agent, the estimated outflow rates lead to accretion rates of
$1.2\times10^{-3}M_{\odot}\,yr^{-1}$ and $4.2\times10^{-2}M_{\odot}\,yr^{-1}$ respectively for the near and far
distances for I18264 and $1.9\times10^{-5}M_{\odot}\,yr^{-1}$ for I23151. Taking into account the missing flux due
to the the missing short spacings for I23151, the accretion rate in this region should be higher. Thus for both
sources the accretion rates are high enough to overcome the radiation pressure from the central objects and form
massive stars.

To summarize, our molecular outflow data derived from high resolution observations toward two luminous sources
show morphologies and kinematics similar to those of the low-mass sources. The outflows can be interpreted by jet
or wide-angle wind entrainment models. The presented data and analysis further support that massive stars up to
$10^5 L_{\odot}$ ($\sim30 M_{\odot}$) form via disk mediated accretion processes as low-mass stars.

\acknowledgments{We are grateful to F. Geuth and the staff at PdBI for their help in the observations and data
reduction. K. Q. acknowledges the support of the Grant 10128306 from NSFC. H. B. acknowledges financial support by
the Emmy-Noether-Program of the Deutsche Forschungsgemeinschaft (DFG, grant BE2578).}

\clearpage

\clearpage

\clearpage

\clearpage

\clearpage

\clearpage

\clearpage

\clearpage

\clearpage

\clearpage

\clearpage

\clearpage

\begin{deluxetable}{ccccccccc}
\tablewidth{0pc} \tabletypesize{\scriptsize} \tablecaption{Distances and derived physical parameters for the two
sources \label{table1}} \tablehead{ \colhead{$source$} & \colhead{$distance$} &
\colhead{$M_{core}$\tablenotemark{a}} & \colhead{$M_{dense}$\tablenotemark{b}}
&\colhead{$M_{blue}$\tablenotemark{c}} & \colhead{$M_{red}$\tablenotemark{d}} &
\colhead{$M_{out}$\tablenotemark{e}} &
\colhead{$\dot{M}_{out}$\tablenotemark{f}} & \colhead{$t_{dyn}$\tablenotemark{g}} \\
\colhead{} & \colhead{(kpc)} & \colhead{($M_{\odot}$)} & \colhead{($M_{\odot}$)} & \colhead{($M_{\odot}$)} &
\colhead{($M_{\odot}$)} & \colhead{($M_{\odot}$)} & \colhead{($10^{-4}\,M_{\odot}\,yr^{-1}$)} &
\colhead{($10^{4}\,yr$)} }
\startdata  I18264   & 3.5  & 570  & 3900  & 3.2 & 17  & 20.2 & 34  & 0.5  \\
                     & 12.5 & 7300 & 50000 & 41  & 220 & 261  & 120 & 1.8  \\
            I23151   & 5.7  & 170  & 860   & 2.2 & 0.5 & 2.7  & 2.3 & 1.2  \\
\enddata
\tablenotetext{a}{masses of the dust-gas cores derived from the 3.4 mm continuum emission}
\tablenotetext{b}{masses of the dense ambient gas derived from the H$^{13}$CO$^+$ emission}
\tablenotetext{c}{outflow masses derived from the blueshifted SiO emission} \tablenotetext{d}{outflow masses
derived from the redshifted SiO emission} \tablenotetext{e}{total outflow masses derived from the sum of the bule-
and redshifted outflow masses} \tablenotetext{f}{mass outflow rates derived from $M_{red}$ for I18264 and from
$M_{out}$ for I23151} \tablenotetext{g}{dynamical time scale}
\end{deluxetable}

\clearpage

\begin{deluxetable}{ccccccc}
\tablewidth{0pc} \tabletypesize{\scriptsize} \tablecaption{Physical parameters derived from the ammonia emission
for I18264. \label{table2}} \tablehead{ \colhead{position\tablenotemark{a}} & \colhead{$\tau(1,1,m)$} &
\colhead{$T_{rot}(2,2:1,1)$} & \colhead{$T_{ex}$} &
\colhead{$T_{kin}$} & \colhead{$n(H_2)$} & \colhead{${\Delta}v$}\\
\colhead{} & \colhead{} & \colhead{(K)} & \colhead{(K)} & \colhead{(K)} & \colhead{($10^5$ cm$^{-3}$)} &
\colhead{(kms$^{-1})$} }
\startdata  left   &  4.0  &  24  & 23 &  30 & 1.6 & 3.1 \\
            middle &  2.7  &  30  & 25 &  45 & 1.0 & 1.9 \\
            right  &  2.7  &  24  & 24 &  32 & 1.6 & 2.5 \\
            mean\tablenotemark{b}   & 3.1 & 26 & 24 &  36 & 1.4 & 2.5 \\
\enddata
\tablenotetext{a}{marked as pluses in the left panel of Fig. \ref{nh3}} \tablenotetext{b}{averaged from the three
positions}
\end{deluxetable}

\clearpage

\end{document}